\documentclass[prb,amsmath,amssymb,superscriptaddress,twocolumn]{revtex4}
\usepackage{amsmath}
\usepackage{braket}
\usepackage{graphicx}
\usepackage{multirow}
\usepackage{bbm,color,ulem}
\allowdisplaybreaks
\begin{document}
\title{Failure of Kohn's theorem and the apparent failure of the $f$-sum rule in intrinsic Dirac-Weyl materials in the presence of a filled Fermi sea}
\author{Robert E.\ Throckmorton}
\author{S.\ Das Sarma}
\affiliation{Condensed Matter Theory Center and Joint Quantum Institute, Department of Physics, University of Maryland, College Park, Maryland 20742-4111 USA}
\date{\today}
\begin{abstract}
Kohn's theorem and the $f$-sum rule are powerful theorems, the first applying to translationally
invariant single-band electronic systems with parabolic electronic dispersion relations and the
second applying to materials in general, that impose restrictions on the effects of electron-electron
interactions on electrical conductivity and on dielectric response, respectively.  We show rigorously
that Kohn's theorem does {\it not} hold for intrinsic Dirac-Weyl materials with filled Fermi seas
where the chemical potential is pinned at the band touching points.  We also demonstrate that
the low-energy effective ``relativistic'' theories used in many-body studies of these materials
violate the $f$-sum rule due to the neglect of the full band structure of the materials in the
effective low-energy relativistic approximations.
\end{abstract}
\maketitle

\section{Introduction}
Kohn's theorem\cite{KohnPR1961} and the $f$-sum rule\cite{NozieresPR1957} are very powerful theorems
regarding the effects of electron-electron interactions in electron systems.  Kohn's theorem states
that interactions do not change the cyclotron resonance frequencies in single-band metals with parabolic band dispersion
relations strictly obeying translational invariance or, if applied in the absence of a magnetic
field, that electron-electron interactions cannot degrade the current in the system and therefore
cannot affect the dc long-wavelength electrical conductivity.  As a result, in a clean material with a
parabolic dispersion relation, the conductivity cannot be changed except by umklapp scattering (which,
arising from an underlying lattice, explicitly breaks translational invariance and momentum conservation) or
Baber scattering (which involves multiband systems also manifesting a breaking of translational invariance).
The $f$-sum rule is a restriction on the dielectric response of materials in general and is as follows.  If
we let $\epsilon(\vec{q},\omega)$ be the dielectric function, then it must satisfy the identity
\begin{equation}
\int_{0}^{\infty}d\omega\,\omega\,\mbox{Im }\left [\frac{1}{\epsilon(\vec{q},\omega)}\right ]=-\frac{2\pi^2 e^2 n}{m}, \label{Eq:fSumRule}
\end{equation}
where $e$ is (the absolute value of) the electron charge, $m$ is the noninteracting mass of the charge
carriers, and $n$ is their number density.  This exact formula applies specifically to three-dimensional
materials, but the invariance of this $f$ sum when interactions are introduced will extend to other dimensions
as well. The right-hand side is also equal to $-\frac{\pi}{2}\omega_p^2$, where $\omega_p$ is the plasma
frequency (which depends on the wave number in two- and one-dimensional systems). We note that the right-hand
side of Eq.\ \eqref{Eq:fSumRule} depends only on the density of carriers and the band mass and thus must remain
constant, even in the presence of interactions.  Thus, Kohn's theorem is a statement of the inability
of electron-electron interactions to affect the long-wavelength dc conductivity of a parabolically
dispersing electron system, while the $f$-sum rule is a similar statement that such interactions
cannot change the long-wavelength plasma frequency in {\it any} material.  We note that the effective metal
in the case of Kohn's theorem has a single band with a partially filled Fermi surface containing $n$
carriers (electrons or holes) per unit volume.  The current work, by contrast, focuses on systems with a
filled Fermi sea and an empty Fermi sea with the chemical potential pinned at the band touching point in between,
as, for example, in Dirac-Weyl materials.  In addition (and this is a crucial pint in our theory), the
noninteracting band dispersion in the system is approximated by the linear chiral relativistic massless
description appropriate at low energies.  Such a relativistic Dirac description for graphene is widespread
in the literature.

We note two facts that allow us to see where these theorems originate.  In the case of Kohn's theorem, it
is related to the fact that, for particles with a parabolic dispersion relation in translationally invariant
systems, momentum and velocity are proportional to each other, so that the total current, which is proportional
to the sum of the velocities of the electrons, is necessarily proportional to the total momentum.  As a result,
we would expect that electron-electron interactions, by virtue of momentum conservation, must not alter the
current; that is, momentum conservation automatically implies velocity or current conservation.  Similar classical
arguments about the inequivalence of momentum and velocity conservation have been presented previously in the
literature for graphene\cite{RoldanPRB2010,SunNJP2012}.  The same intuition, however, does not apply to Dirac
electrons with linear dispersion.  For one, there is no direct proportionality between momentum and velocity---all
electrons travel at the same speed, regardless of momentum.  This opens up the possibility that electron-electron
interactions can, in fact, degrade the current in such a system since momentum conservation does not automatically
imply velocity conservation.  This line of argument, of course, is purely classical in nature; we will put it on
a more rigorous and quantum-mechanical footing in this work.

As for the $f$-sum rule, it emerges from evaluating the double commutator $[[H,\rho(\vec{q})],\rho(-\vec{q})]$,
where $H$ is the Hamiltonian of the system, now that of any material, and $\rho(\vec{q})$ is the density operator
in momentum space.  The integral on the left-hand side of the $f$-sum rule is proportional to the expectation value
of this double commutator.  It turns out that it is proportional only to $q^2$ and to the total electron number
operator.  This means that the expectation value, and thus the $f$ sum, is invariant under the introduction of
anything that changes only the energy distribution of the electrons, such as finite temperature or interactions.
We will, however, find that the double commutator that yields the $f$-sum rule, when evaluated for the low-energy
effective theories describing intrinsic Dirac-Weyl materials, cannot be expressed in terms of the total electron
number operator; we instead obtain an operator whose expectation value depends on the details of the electrons'
energy distribution, and thus can be affected by interactions.  This is, of course, a direct consequence of the widely
used effective Dirac-Weyl relativistic approximation for the system.

We mention that the fact that in Dirac systems electron-electron interactions could affect the electrical
conductivity was previously discussed in the literature in the context of quantum criticality associated
with the Dirac point\cite{FritzPRB2008}, and as such, the violation of Kohn's Theorem in the Dirac system is
implicitly known in the literature.  Our work, however, puts the violation of Kohn's Theorem on firm formal
footing by considering the Dirac system. Our formal proof follows Kohn's derivation and shows that the theorem
is violated in Dirac systems without any reference to quantum criticality.  Also, it is well known that in
the presence of the breaking of translational invariance by some mechanism, electron-electron interactions
can, in fact, affect the electrical conductivity even in a system with parabolic energy dispersion (but the
interaction effect must disappear as the translational invariance is restored).  Some well-known examples
of such translational-invariance-breaking mechanisms leading to the violation of Kohn's Theorem in its pristine
form even for nominally parabolic band systems are umklapp scattering in lattice systems, Baber scattering
in transition metals (or generally in two-band systems)\cite{BaberPRSA1937}, electron-hole scattering in
semiconductors (or generally in any multicomponent system with more than one carrier species differing in
charge or mass), electronic screening of impurity disorder\cite{DasSarmaPRL1999}, hydrodynamic effects associated
with strong interaction\cite{LucasArXiv} and Altshuler-Aronov-type interaction corrections in the presence of
diffusive carriers\cite{AltshulerSSC1983}.  The interesting physics in Dirac materials is that interaction
affects the conductivity intrinsically by virtue of the fundamental breaking of Kohn's theorem as we show in
this paper---no explicit mechanism breaking translational invariance (such as umklapp or Baber scattering or
disorder, etc.) is necessary.  Of course, one could argue that in a solid-state system a linear band dispersion
cannot arise unless the translational invariance is somehow broken (i.e., a lattice must somehow be present),
but the standard continuum theory we use in the current work does not make any explicit reference to an explicit
momentum-conservation-breaking mechanism from outside.

The same applies to the apparent breakdown of the $f$-sum rule which in ordinary materials arises from number
conservation and thus appears to be sacrosanct.  However, in low-energy effective theories of Dirac materials,
the gapless nature of the electron and hole bands touching at the Dirac point leads to an apparent breakdown of
number conservation since the infinitely filled valence band provides a mechanism for zero-energy excitations
violating the $f$-sum rule in its pristine form, as we show in our work.  We should reiterate here that our
statements about violations of the $f$-sum rule apply to only the low-energy effective theories of materials
with Dirac points, not to putative exact models of such materials, which {\it must} obey the $f$-sum rule and,
as a matter of fact, do not possess an infinite negative-energy sea of electronic modes as the low-energy theory
does.  Thus, the $f$-sum rule must be restored by virtue of the number conservation in a full band description of the system,
but it is violated in the low-energy Dirac description, as we show explicitly in the current paper.  We mention here
that the fact that the na\"ive $f$-sum rule is violated in various low-energy effective model Hamiltonians has been
demonstrated in the literature in various contexts\cite{GoldmanPRA1982,CenniNPA2001,LimtragoolPRB2017}, and our
current work focuses explicitly on the low-energy models of graphenelike systems of great current interest,
showing explicitly that the low-energy effective Hamiltonians here indeed violate the na\"ive $f$-sum rule defined
by Eq.\ \eqref{Eq:fSumRule}.

While all of our discussion so far has been about Dirac materials with linear dispersion, we should
emphasize that the basic physics described above is by no means limited to just Dirac materials with linear
dispersion---any material featuring two touching bands will display these effects as well, as long as the chemical
potential is pinned at the band touching point (i.e., the system is undoped).  In particular, the breakdown of 
Kohn's theorem and the apparent breakdown of the $f$-sum rule also occur in materials with quadratic band touching
points (e.g., bilayer graphene), which we also explicitly and formally show in this work.  The key for the
breakdown is not the dispersion linearity, but the simultaneous presence of a filled Fermi sea and an empty Fermi
sea in the system with the chemical potential being precisely at the band touching point.  Thus, all intrinsic
Dirac-Weyl-type materials in all dimensions violate Kohn's theorem and appear to violate the $f$-sum rule if a
low-energy effective theory is employed if the chemical potential is at the Dirac point.  Again, the $f$-sum rule
will be restored in a full band description of these systems, but not in the effective low-energy models used
extensively in the literature.

We also emphasize that an exact theory starting from the lattice model of Dirac materials and including all bands and
valleys will, by definition, obey the $f$-sum rule since the $f$-sum rule is simply a restatement of the particle
conservation law.  But such a theory will have to be completely numerical from the beginning.  What we consider in
the current work is the effective low-energy two-band continuum Dirac description in which the system is a semimetal
described by only linear or quadratic and chiral particle-hole bands.  We show that such a low-energy continuum theory violates
the $f$-sum rule explicitly, and therefore, analytical theories using the low-energy continuum description should
take into account this apparent violation of the $f$-sum rule in the dielectric response.  In a purely numerical
exact theory, such a failure plays no role at all.

The rest of this work is organized as follows.  We dedicate Secs.\ II and III to reviewing the proofs of Kohn's
theorem for partially filled single-band systems with parabolic dispersions and of the $f$-sum rule for general
materials, respectively, and then showing that Kohn's theorem breaks down for systems with massless Dirac
dispersions or quadratic band touching points and that the commonly used low-energy effective theories for these
materials violate the $f$-sum rule.  We then provide conclusions in Sec.\ IV.

\section{Kohn's Theorem}

\subsection{Dirac electrons}
We begin by addressing whether or not a theorem similar to Kohn's theorem holds for Dirac electrons.
We will show that, in fact, this theorem is dependent on the parabolic dispersion and a finite Fermi
surface by showing that it fails for massless Dirac electrons. The essential difference is as follows.
For electrons with a parabolic dispersion, the total momentum, the sum of the velocities of the electrons,
and the current are all proportional to each other, and thus, conservation of momentum implies conservation
of the other two quantities. This is not true for massless Dirac electrons, however---the speed of the
electrons is always the same, no matter what their momenta may be.  Thus, velocity conservation fails
explicitly even when momentum is conserved in the continuum model.

It might be helpful to first review Kohn's original proof, as we will model our calculation on it.  We
start with the Hamiltonian,
\begin{equation}
H=\sum_{j=1}^{N}\frac{P_j^2}{2m}+\sum_{1\leq j<k\leq N}u(\vec{r}_j-\vec{r}_k), \label{Eq:Hamiltonian}
\end{equation}
where $u(\vec{r}_j-\vec{r}_k)$ is a two-body interaction between electrons $j$ and $k$ and $\vec{P}_j=\vec{p}_j+\frac{e}{c}\vec{A}(\vec{r}_j)$
is the kinetic momentum of electron $j$.  We assume that $u(-\vec{r})=u(\vec{r})$ and work in the Landau
gauge, so that $\vec{P}_j=\vec{p}_j+\frac{eB}{c}x_j\hat{\vec{y}}$.

Working in the Heisenberg picture, the equation of motion for the total (kinetic) momentum of the system,
$\vec{P}=\sum_{j}P_j$, is
\begin{equation}
\frac{d\vec{P}}{dt}=\frac{i}{\hbar}[H,\vec{P}]=-\frac{e}{mc}\vec{P}\times\vec{B}.
\end{equation}
At this point, we can simply set $\vec{B}=0$ to obtain the familiar statement of conservation of
total momentum and thus, as stated earlier, current.  If, however, we retain the magnetic field,
then we find that
\begin{equation}
[H,P_{\pm}]=\pm\hbar\omega_c P_{\pm},
\end{equation}
where $P_{\pm}=P_x\pm iP_y$ and $\omega_c=\frac{eB}{mc}$ is the cyclotron frequency.  This means that
the $P_{\pm}$ operators act like the raising and lowering operators for the harmonic oscillator---applied
to an eigenstate of $H$ with energy $E$, they produce another eigenstate with an energy $E\pm\hbar\omega_c$.
Therefore, the cyclotron resonance frequencies are unchanged.

Although Kohn's original proof is only for explicitly translationally invariant systems as in Eq.\ \eqref{Eq:Hamiltonian},
where there is no external one-particle potential destroying translational invariance (in fact,
the theorem fails generically in the presence of such a one-electron spatially varying potential since
the center-of-mass and relative coordinates are no longer separable, and hence interaction effects can
then modify the center-of-mass motion), it is possible to generalize Kohn's theorem to a situation where
an explicit parabolic external potential is applied to the system\cite{BreyPRB1989,LiPRB1991}.  This is
simply because the parabolic potential allows the separation of the center-of-mass and relative coordinates,
thus preserving Kohn's theorem in spite of an apparent breaking of the translational invariance.

Now we turn our attention to the case of a massless Dirac fermion and attempt to replicate the above proof
for this case.  The Hamiltonian is
\begin{equation}
H=v_F\sum_{j=1}^N\vec{\sigma}_j\cdot\vec{P}_j+\sum_{1\leq j<k\leq N}u(\vec{r}_j-\vec{r}_k),
\end{equation}
where $\vec{\sigma}_j$ is the vector of Pauli matrices acting on electron $j$.  The physical meaning of the
Pauli matrices will depend on the material under consideration; as an example, in graphene, this represents
the sublattice that the electron is on.  Here, we consider a system with just one flavor of Dirac
fermion for simplicity; additional flavors will not affect our conclusions.  The velocity operator is
given by
\begin{equation}
\vec{v}_j=\frac{i}{\hbar}[H,\vec{r}_j]=v_F\vec{\sigma}_j.
\end{equation}
Note that this is {\it not} proportional to the momentum operator; as implied earlier, this is the key
fact that will lead to the failure of Kohn's theorem for this case.

We begin by determining the equation of motion for the kinetic momentum.  The commutator of a component of
the kinetic momentum $P_{k,\alpha}$ of electron $k$ is
\begin{eqnarray}
[H,P_{k,\alpha}]&=&-i\frac{\hbar eB}{c}v_{k,x}\delta_{\alpha,y}+i\frac{\hbar eB}{c}v_{k,y}\delta_{\alpha,x} \cr
&-&i\hbar\sum_{j=k+1}^N\frac{\partial u(\vec{r}_k-\vec{r}_j)}{\partial x_{k,\alpha}}+i\hbar\sum_{j=1}^{k-1}\frac{\partial u(\vec{r}_j-\vec{r}_k)}{\partial x_{k,\alpha}}. \nonumber \\
\end{eqnarray}
The commutator with the total momentum $P_{\alpha}=\sum_{j=1}^N P_{j,\alpha}$ is then
\begin{equation}
[H,P_{\alpha}]=-i\frac{\hbar eB}{c}(v_x\delta_{\alpha,y}-v_y\delta_{\alpha,x}).
\end{equation}
We may write this equation for the $x$ and $y$ components in a different way; defining $P_{\pm}=P_x\pm iP_y$
as before, we find that
\begin{equation}
[H,P_{\pm}]=\pm\frac{\hbar eB}{c}v_{\pm},
\end{equation}
where $v_{\alpha}=\sum_{j=1}^N v_{j,\alpha}$ is the sum of the electrons' velocities, which is proportional to
the total current, and $v_{\pm}=v_x\pm iv_y$.  Therefore,
\begin{equation}
\frac{dP_{\pm}}{dt}=\pm i\frac{eB}{c}v_{\pm}.
\end{equation}
Because the velocity operator is not proportional to the momentum operator, we cannot necessarily
conclude from this result that, in the absence of a magnetic field, the sum of the electrons' velocities,
and thus the total current, will be conserved.  In fact, the above result also demonstrates explicitly
that the cyclotron resonance frequency in a Dirac material will, indeed, be renormalized by electron-electron
interaction in direct violation of Kohn's theorem.  Indeed, the dependence of the cyclotron resonance on
interaction effects has been experimentally observed in graphene, which is the prototypical Dirac material\cite{HenriksenPRL2010}.

Next, we determine the equation of motion for the sum of the electron velocity operators.  The commutator
of $v_{\pm}$ with the Hamiltonian is
\begin{equation}
[H,v_{\pm}]=\pm v_F\sum_{k=1}^N(v_{k,\pm}P_{k,z}-v_{k,z}P_{k,\pm}),
\end{equation}
so that the corresponding equation of motion is
\begin{equation}
\frac{dv_{\pm}}{dt}=\pm\frac{iv_F}{\hbar}\sum_{k=1}^N(v_{k,\pm}P_{k,z}-v_{k,z}P_{k,\pm}).
\end{equation}
We thus have obtained an expression that cannot be written entirely in terms of sums of the electrons'
momenta and velocities, although at this stage we cannot explicitly see any dependence of the equations
of motion for the sum of the velocities on the electron-electron interaction.  We now consider the equation
of motion for $\sum_{k=1}^N(v_{k,\pm}P_{k,z}-v_{k,z}P_{k,\pm})$. To this end, we will need the identity,
$[A,BC]=[A,B]C+B[A,C]$, along with the following commutators:
\begin{eqnarray}
[H,P_{k,z}]&=&-i\hbar\sum_{j=k+1}^N\frac{\partial u(\vec{r}_k-\vec{r}_j)}{\partial z_k} \cr
&+&i\hbar\sum_{j=1}^{k-1}\frac{\partial u(\vec{r}_j-\vec{r}_k)}{\partial z_k}, \\
\left [H,v_{k,z}\right ]&=&\tfrac{1}{2}v_F(v_{k,-}P_{k,+}-v_{k,+}P_{k,-}).
\end{eqnarray}
Combining these results, we obtain the equation of motion,
\begin{eqnarray}
&&\frac{d}{dt}\sum_{j=1}^N(v_{k,\pm}P_{k,z}-v_{k,z}P_{k,\pm}) \cr
&=&\sum_{j=1}^N\left\{\frac{iv_F}{\hbar}(v_{j,\pm}P_{j,z}-v_{j,z}P_{j,\pm})P_{j,z}\right. \cr
&-&\frac{iv_F}{2\hbar}(P_{j,+}v_{j,-}-P_{j,-}v_{k,+})P_{j,\pm} \cr
&-&v_{j,\pm}\left (\sum_{k=1}^{j-1}\frac{\partial u(\vec{r}_k-\vec{r}_j)}{\partial z_j}-\sum_{k=j+1}^N\frac{\partial u(\vec{r}_j-\vec{r}_k)}{\partial z_j}\right ) \cr
&-&v_{j,z}\left [\pm\frac{ieB}{c}v_{j,\pm}-\sum_{k=1}^{j-1}\left (\frac{\partial u(\vec{r}_k-\vec{r}_j)}{\partial x_j}\pm i\frac{\partial u(\vec{r}_k-\vec{r}_j)}{\partial y_j}\right )\right. \cr
&+&\left.\left.\sum_{k=j+1}^N\left (\frac{\partial u(\vec{r}_j-\vec{r}_k)}{\partial x_j}\pm i\frac{\partial u(\vec{r}_j-\vec{r}_k)}{\partial y_j}\right )\right ]\right\}.
\end{eqnarray}

Similarly, we find that the equation of motion for $v_z$ is
\begin{equation}
\frac{dv_z}{dt}=\frac{iv_F}{2\hbar}\sum_{k=1}^N(v_{k,-}P_{k,+}-v_{k,+}P_{k,-}),
\end{equation}
and the equation of motion of the operator on the right-hand side of it is
\begin{eqnarray}
&&\frac{d}{dt}\sum_{j=1}^N(v_{j,-}P_{j,+}-v_{j,+}P_{j,-}) \cr
&=&\sum_{j=1}^N\left\{-\frac{iv_F}{\hbar}(v_{j,+}P_{j,z}-v_{j,z}P_{j,+})P_{j,-}\right. \cr
&-&\frac{iv_F}{\hbar}(v_{j,-}P_{j,z}-v_{j,z}P_{j,-})P_{j,+} \cr
&+&v_{j,-}\left [\frac{ieB}{c}v_{j,+}-\sum_{k=1}^{j-1}\left (\frac{\partial u(\vec{r}_j-\vec{r}_k)}{\partial x_{k}}+i\frac{\partial u(\vec{r}_j-\vec{r}_k)}{\partial y_{k}}\right )\right. \cr
&+&\left. \sum_{k=j+1}^N\left (\frac{\partial u(\vec{r}_k-\vec{r}_j)}{\partial x_{k}}+i\frac{\partial u(\vec{r}_k-\vec{r}_j)}{\partial y_{k}}\right )\right ] \cr
&-&v_{j,+}\left [-\frac{ieB}{c}v_{j,-}-\sum_{k=1}^{j-1}\left (\frac{\partial u(\vec{r}_j-\vec{r}_k)}{\partial x_{k}}-i\frac{\partial u(\vec{r}_j-\vec{r}_k)}{\partial y_{k}}\right )\right. \cr
&+&\left.\left.\sum_{k=j+1}^N\left (\frac{\partial u(\vec{r}_k-\vec{r}_j)}{\partial x_{k}}-i\frac{\partial u(\vec{r}_k-\vec{r}_j)}{\partial y_{k}}\right )\right ]\right\}.
\end{eqnarray}
The terms that depend on the electron-electron interaction do not cancel out of these expressions due to the
prefactors of $v_{j,\pm}$ and $v_{j,z}$, and thus, we conclude that the time evolution of the sum of the electrons'
velocities, and thus the total current in the system, depends on the interaction, in contrast to the case with
a parabolic dispersion.

We emphasize that our formal results arise from the fact that the relative and center-of-mass coordinates are
hopelessly intermixed in the dynamics of Dirac systems, and thus, electron-electron interaction, in spite of
being dependent on only relative coordinates, affects the total momentum, and hence the conductivity. The fact
that the correct calculation of the dielectric function for the Dirac problem must account for the intraband
and interband processes on an equal footing was already pointed out in Ref.\ \onlinecite{HwangPRB2007}.

To help explicitly illustrate the basic physics behind Kohn's theorem for electrons with parabolic dispersions and its
failure for Dirac electrons, it might be helpful to consider collisions between two electrons in single-particle non-interacting
eigenstates in each case.  We may equivalently consider what follows to be a (semi-)classical argument.  This will also
help us to quantify how severe the effects of this breakdown will be for Dirac electrons.  In both cases, the conservation
of momentum is very well known:
\begin{equation}
\vec{p}_{1,i}+\vec{p}_{2,i}=\vec{p}_{1,f}+\vec{p}_{2,f}.
\end{equation}
In the case of a parabolic dispersion, the momentum is just $\vec{p}=m\vec{v}$, so that, dividing by $m$,
\begin{equation}
\vec{v}_{1,i}+\vec{v}_{2,i}=\vec{v}_{1,f}+\vec{v}_{2,f}.
\end{equation}
Therefore, the sum of the velocities of the two electrons is also conserved, and thus, this collision will not degrade the
total current carried by them.  On the other hand, if the electrons have a Dirac dispersion, then the velocity of an
electron
\begin{equation}
v_j=\frac{\partial E}{\partial p_j}=\frac{v_F}{|\vec{p}|}p_j.
\end{equation}
In this case, the conservation of momentum becomes
\begin{equation}
\vec{v}_{1,i}|\vec{p}_{1,i}|+\vec{v}_{2,i}|\vec{p}_{2,i}|=\vec{v}_{1,f}|\vec{p}_{1,f}|+\vec{v}_{2,f}|\vec{p}_{2,f}|.
\end{equation}
In contrast to the parabolic case, this equation does {\it not} necessarily imply conservation of the total velocity.
Thus, electron-electron interaction, in spite of being momentum conserving, can indeed relax the charge current flow
and hence lead to a finite electrical conductivity.  However, total velocity conservation is approximately true in the
case of low-energy excitations if there is a finite chemical potential, i.e., all electronic wave vectors are near the
Fermi surface.  In this case, $p_{n,i/f}\approx\hbar k_F$.  This approximation should hold very well in cases where the
temperature is far below the chemical potential, i.e., $k_BT\ll\mu$, since the Pauli exclusion principle will help to
``freeze out'' scattering into or from states with momenta other than $\hbar k_F$.  Note, however, that at the Dirac
point, where $\mu=0$ in our notation, the system is {\it always} susceptible to interaction effects and is thus a
non-Fermi liquid, which has led people to dub the Dirac point a quantum critical point lying between an electron metal
and a hole metal.  Thus, for an undoped intrinsic system with the chemical potential at the Dirac point, the system
always violates current conservation in the presence of electron-electron scattering.

\subsection{Quadratic band touching}
We now turn our attention to the case of a quadratic band touching point, e.g., in bilayer graphene.  We will find
that similar effects occur here as well provided that the Fermi level is pinned at the band touching point
(i.e., no partial band filling).  The Hamiltonian for this case is
\begin{eqnarray}
H&=&\frac{1}{2m}\sum_{j=1}^N [(P_{j,x}^2-P_{j,y}^2)\sigma_{j,x}+2P_{j,x}P_{j,y}\sigma_{j,y}] \cr
&+&\sum_{1\leq j<k\leq N}u(\vec{r}_j-\vec{r}_k),
\end{eqnarray}
where all symbols have the same basic meaning as before.  If we determine the velocity operators for this system,
we get
\begin{eqnarray}
v_{j,x}=\frac{P_{j,x}}{m}\sigma_{j,x}+\frac{P_{j,y}}{m}\sigma_{j,y}, \\
v_{j,y}=\frac{P_{j,x}}{m}\sigma_{j,y}-\frac{P_{j,y}}{m}\sigma_{j,x}.
\end{eqnarray}
We note that the direct proportionality between momentum and velocity is broken in a different way here---the
velocity component operators depend on {\it both} components of the (kinetic) momentum operator.  We should
emphasize here, however, that this lack of direct proportionality between the velocity and momentum operators
is just a symptom of the underlying physics at work here, namely, the presence of a filled negative-energy Fermi
sea.

If we now determine the equations of motion for the kinetic momentum, we obtain similar results as before.
Letting $P_\alpha=\sum_{j}P_{j,\alpha}$ be the total momentum, we find that
\begin{eqnarray}
\frac{dP_x}{dt}=-\frac{eB}{c}v_y, \\
\frac{dP_y}{dt}=\frac{eB}{c}v_x.
\end{eqnarray}
We now want to determine the equations of motion for the velocity.  Following a procedure similar to that for the
Dirac case, we find that
\begin{eqnarray}
\frac{dv_{k,x}}{dt}&=&-\frac{eB}{mc}v_{k,y}\sigma_{k,x}+\frac{eB}{mc}v_{k,x}\sigma_{k,y} \cr
&-&\frac{1}{m}\left [\sum_{j=1}^{k-1}\frac{\partial u(\vec{r}_j-\vec{r}_k)}{\partial x_k}-\sum_{j=k+1}^{N}\frac{\partial u(\vec{r}_k-\vec{r}_j)}{\partial x_k}\right ]\sigma_{k,x} \cr
&-&\frac{1}{m}\left [\sum_{j=1}^{k-1}\frac{\partial u(\vec{r}_j-\vec{r}_k)}{\partial y_k}-\sum_{j=k+1}^{N}\frac{\partial u(\vec{r}_k-\vec{r}_j)}{\partial y_k}\right ]\sigma_{k,y} \cr
&+&\frac{1}{\hbar}\frac{P_{k,x}}{m}\frac{P_{k,x}P_{k,y}}{m}\sigma_{k,z}-\frac{1}{\hbar}\frac{P_{k,y}}{m}\frac{P_{k,x}^2-P_{k,y}^2}{2m}\sigma_{k,z} \nonumber \\
\end{eqnarray}
and
\begin{eqnarray}
\frac{dv_{k,y}}{dt}&=&-\frac{eB}{mc}v_{k,x}\sigma_{k,x}-\frac{eB}{mc}v_{k,y}\sigma_{k,y} \cr
&-&\frac{1}{m}\left [\sum_{j=1}^{k-1}\frac{\partial u(\vec{r}_j-\vec{r}_k)}{\partial x_k}-\sum_{j=k+1}^{N}\frac{\partial u(\vec{r}_k-\vec{r}_j)}{\partial x_k}\right ]\sigma_{k,y} \cr
&+&\frac{1}{m}\left [\sum_{j=1}^{k-1}\frac{\partial u(\vec{r}_j-\vec{r}_k)}{\partial y_k}-\sum_{j=k+1}^{N}\frac{\partial u(\vec{r}_k-\vec{r}_j)}{\partial y_k}\right ]\sigma_{k,x} \cr
&-&\frac{1}{\hbar}\frac{P_{k,x}}{m}\frac{P_{k,x}^2-P_{k,y}^2}{2m}\sigma_{k,z}-\frac{1}{\hbar}\frac{P_{k,y}}{m}\frac{P_{k,x}P_{k,y}}{m}\sigma_{k,z}. \nonumber \\
\end{eqnarray}
We see that, if we were to sum the velocities of all of the particles, then the terms involving the interaction
would not cancel out due to the $\sigma_{k,\alpha}$ factors.  As a result, we find that Kohn's theorem is also
broken for a system with quadratic band touching points.  As noted before, the fact that the exact nature of the
dispersion does not matter points to the underlying reason for this failure of Kohn's theorem being the presence
of an empty electron band and a filled hole band with the Fermi level exactly at the band touching point, rather
than just a partially filled electron band as in ordinary metals.  Again, Kohn's theorem will apply when the Fermi
energy (in a doped system) is high compared with temperature so that interband transitions are no longer important.

\section{$f$-sum rule}
We now turn our attention to the $f$-sum rule, also referred to as the Thomas-Reiche-Kuhn sum rule for metals.
In exact models of general material systems, it has been shown that what is known as the $f$-sum rule, Eq.\ \eqref{Eq:fSumRule},
holds with the frequency or energy integral in Eq.\ \eqref{Eq:fSumRule} going over all energies and hence all
bands.  None of the quantities on the right-hand side of this relation are dependent on such things as temperature
or the presence of interactions.  Therefore, this relation provides a powerful constraint on models of the dielectric
function for materials, as it can be calculated for noninteracting electrons, but must be followed even in the
presence of interactions.  In fact, this sum rule necessarily restricts the high-frequency long-wavelength dielectric
function of a simple metal to the form
\begin{equation}
\epsilon(\vec{q},\omega)=1-\frac{\omega_p^2}{\omega^2},
\end{equation}
where $\omega_p$ is the standard electronic plasma frequency.

We will now illustrate that the $f$-sum rule is violated by the low-energy effective theories of materials with
massless Dirac points often used in many-body calculations---the right-hand side of the analogous relation will
actually depend on the details of the energy distribution of the electrons, and thus interactions or even temperature
can change it.  This failure of the simple $f$-sum rule arises from the presence of an infinite filled negative-energy
hole Fermi sea in these effective models.  In a complete model of the system with all bands, the $f$-sum rule is
an identity which must always be obeyed.

We will begin by stating the most general form of the $f$-sum relation.  The right-hand side is proportional
to the expectation value of the double commutator, $[[H,\rho(\vec{q})],\rho(-\vec{q})]$, where $H$ is the
Hamiltonian and $\rho(\vec{q})$ is the density operator in momentum space.  Following a derivation similar to
that of Ref.\ \onlinecite{MahanBook}, we find that
\begin{equation}
\int_0^{\infty}d\omega\,\omega\,\mbox{Im }\left [\frac{1}{\epsilon(\vec{q},\omega)}\right ]=\frac{\pi e^2V_C(\vec{q})}{2\hbar^2 V}\braket{[[H,\rho(\vec{q})],\rho(-\vec{q})]},
\end{equation}
where $V$ is the volume of the system and $V_C(\vec{q})$ is the Coulomb interaction in momentum space.  In the
case of an exact Hamiltonian describing a material,
\begin{equation}
H=\sum_{j=1}^N\left [\frac{p_j^2}{2m}+U(\vec{r}_j)\right ],
\end{equation}
where $U(\vec{r})$ is the potential that the electrons are subject to, including the periodic ionic lattice and
disorder, one finds that the double commutator is
\begin{equation}
[[H,\rho(\vec{q})],\rho(-\vec{q})]=-\frac{\hbar^2 q^2}{m}N,
\end{equation}
where $N=\sum_{\vec{k}}n(\vec{k})$ is the total electron number operator.  If we substitute this into the previous
equation and use the fact that, in three dimensions, $V_C(\vec{q})=\frac{4\pi e^2}{q^2}$, we will recover Eq.\ \eqref{Eq:fSumRule}.
The fact that this expression for the double commutator depends on only the total electron number operator is what
leads to the invariance of the $f$ sum under the introduction of interactions.  We show below that this is no longer
true in the extensively used low-energy effective theories in which a relativistic band dispersion is used to describe the
system around the chemical potential.

We now calculate the double commutator for Dirac electrons within the usual low-energy effective theory.  The density
operators commute with the interaction term, so we just need to find it for the noninteracting case.  The noninteracting
part of the Hamiltonian $H_0$ is, in second-quantized form,
\begin{equation}
H_0=\hbar v_F\sum_{\vec{k}}\Psi^{\dag}(\vec{k})\vec{\sigma}\cdot\vec{k}\Psi(\vec{k}),
\end{equation}
where $\Psi^{T}(\vec{k})=[a(\vec{k}),b(\vec{k})]$ is the vector of annihilation operators for pseudospins $a$ and
$b$ (e.g., sublattice), and
\begin{equation}
\rho(\vec{q})=\sum_{\vec{k}}\Psi^{\dag}(\vec{k})\Psi(\vec{k}+\vec{q}).
\end{equation}
Applying the usual anticommutation relations for fermionic operators, we find that
\begin{eqnarray}
[[H,\rho(\vec{q})],\rho(-\vec{q})]&=&\hbar v_F\sum_{\vec{k}}[2\Psi^{\dag}(\vec{k})\vec{\sigma}\cdot\vec{k}\Psi(\vec{k}) \cr
&-&\Psi^{\dag}(\vec{k}+\vec{q})\vec{\sigma}\cdot\vec{k}\Psi(\vec{k}+\vec{q}) \cr
&-&\Psi^{\dag}(\vec{k}-\vec{q})\vec{\sigma}\cdot\vec{k}\Psi(\vec{k}-\vec{q})].
\end{eqnarray}
If we now perform the unitary transformation that diagonalizes $H_0$, we may split this expression into two sets of
terms, which we will denote $C_1$ and $C_2$, i.e., $[[H,\rho(\vec{q})],\rho(-\vec{q})]=C_1+C_2$, where
\begin{eqnarray}
C_1&=&\hbar v_F\sum_{\vec{k}}[2k\Psi^{\dag}_{+}(\vec{k})\Psi_{+}(\vec{k})-2k\Psi^{\dag}_{-}(\vec{k})\Psi_{-}(\vec{k}) \cr
&+&|\vec{k}+\vec{q}|\Psi^{\dag}_{+}(\vec{k}+\vec{q})\Psi_{+}(\vec{k}+\vec{q}) \cr
&-&|\vec{k}+\vec{q}|\Psi^{\dag}_{-}(\vec{k}+\vec{q})\Psi_{-}(\vec{k}+\vec{q}) \cr
&+&|\vec{k}-\vec{q}|\Psi^{\dag}_{+}(\vec{k}-\vec{q})\Psi_{+}(\vec{k}-\vec{q}) \cr
&-&|\vec{k}-\vec{q}|\Psi^{\dag}_{-}(\vec{k}-\vec{q})\Psi_{-}(\vec{k}-\vec{q})]
\end{eqnarray}
and
\begin{eqnarray}
C_2&=&\hbar v_F\sum_{\vec{k}}[\Psi^{\dag}(\vec{k}+\vec{q})\vec{\sigma}\cdot\vec{q}\Psi(\vec{k}+\vec{q}) \cr
&-&\Psi^{\dag}(\vec{k}-\vec{q})\vec{\sigma}\cdot\vec{q}\Psi(\vec{k}-\vec{q})].
\end{eqnarray}
In these expressions, $\Psi_{\pm}(\vec{k})$ is the annihilation operator for electrons in positive- ($+$) and negative-
($-$) energy single-particle eigenstates.  We note that an expression similar to ours was derived for the case of
monolayer graphene in Ref.\ \onlinecite{SabioPRB2008}; there, only the terms corresponding to our $C_2$ are obtained.
It turns out that the $C_1$ terms have a nonzero expectation value; we believe that this arises from the same phenomenon
mentioned therein (the ``anomalous commutator'' problem).

We note that, in contrast to the results obtained from an exact model, the expression that we obtain for $[[H,\rho(\vec{q})],\rho(-\vec{q})]$
from the low-energy effective theory cannot be expressed in terms of the total number of particles; the expectation value
of this expression will necessarily depend on the details of the electron energy distribution.  Therefore, this model for
Dirac electrons violates the $f$-sum rule.

We find similar results for the intrinsic quadratic band touching case as well.  The noninteracting low-energy effective
Hamiltonian, again in second-quantized form, is
\begin{equation}
H_0=\frac{\hbar^2}{2m}\sum_{\vec{k}}\Psi^{\dag}(\vec{k})[(k_x^2-k_y^2)\sigma_x+2k_xk_y\sigma_y]\Psi(\vec{k}),
\end{equation}
where all symbols have meanings similar to those in the Dirac case.  If we now calculate $[[H,\rho(\vec{q})],\rho(-\vec{q})]$,
we find that it is given by $D_1+D_2$, where
\begin{eqnarray}
D_1&=&\frac{\hbar^2}{2m}\sum_{\vec{k}}[2k^2\Psi^{\dag}_{+}(\vec{k})\Psi_{+}(\vec{k})-2k^2\Psi^{\dag}_{-}(\vec{k})\Psi_{-}(\vec{k}) \cr
&-&|\vec{k}+\vec{q}|^2\Psi^{\dag}_{+}(\vec{k}+\vec{q})\Psi_{+}(\vec{k}+\vec{q}) \cr
&+&|\vec{k}+\vec{q}|^2\Psi^{\dag}_{-}(\vec{k}+\vec{q})\Psi_{-}(\vec{k}+\vec{q}) \cr
&-&|\vec{k}-\vec{q}|^2\Psi^{\dag}_{+}(\vec{k}-\vec{q})\Psi_{+}(\vec{k}-\vec{q}) \cr
&+&|\vec{k}-\vec{q}|^2\Psi^{\dag}_{-}(\vec{k}-\vec{q})\Psi_{-}(\vec{k}-\vec{q})],
\end{eqnarray}
\begin{eqnarray}
D_2&=&\frac{\hbar^2}{2m}\sum_{\vec{k}}[\Psi^{\dag}(\vec{k}+\vec{q})M_{+}\Psi(\vec{k}+\vec{q}) \cr
&-&\Psi^{\dag}(\vec{k}-\vec{q})M_{-}\Psi(\vec{k}-\vec{q})],
\end{eqnarray}
and
\begin{widetext}
\begin{equation}
M_{\pm}=\left [\begin{matrix}
0 & (q_x-iq_y)[2k_x\pm q_x-i(2k_y\pm q_y)] \\
(q_x+iq_y)[2k_x\pm q_x+i(2k_y\pm q_y)] & 0
\end{matrix}\right ].
\end{equation}
\end{widetext}
As in the Dirac case, the expectation value of the double commutator will depend on the details of the electronic energy
distribution and thus will be changed by the presence of electron-electron interactions.  Thus, the $f$-sum rule is violated
by low-energy effective field theories of intrinsic undoped Dirac-Weyl systems independent of energy dispersion.

As we already stated in the Introduction, this violation of the $f$-sum rule arises from the presence of the infinite filled valence
band in the continuum Dirac-Weyl system---one can, of course, impose a sum rule by imposing a physical energy or momentum cut off
on the spectrum, but then the result becomes explicitly dependent on this cut off.  Thus, the invariable presence of both intraband
and interband processes and, ultimately, the fact that our low-energy theory ignores such phenomena as deviations from the assumed
linear or quadratic band structure at higher energies and the presence of lower- or higher-energy bands destroy the simplicity of
an $f$-sum rule for these models.

\section{Conclusion}

We have revisited two well-known theorems, Kohn's theorem and the $f$-sum rule, in the context of massless Dirac and Weyl materials.
These theorems are very powerful---Kohn's theorem places restrictions on the ability of electron-electron interactions to affect
cyclotron resonance frequency and dc conductivity for electrons with parabolic dispersions\cite{KohnPR1961}, while the $f$-sum rule
imposes a restriction on the correct mathematical models of dielectric response in any material\cite{NozieresPR1957}.  While the
charge carriers in many semiconductors and metals possess parabolic dispersions, there are a number of materials, including two-dimensional
graphene and three-dimensional Dirac and Weyl materials, that have massless Dirac dispersions instead.  We find that Kohn's theorem breaks down
in such materials and that the low-energy effective theories often employed in studies of many-body effects in massless Dirac and
Weyl materials violate the $f$-sum rule.  Our results imply, for one, that electron-electron interactions can, in fact, change the
electrical conductivity of these materials, even in the absence of umklapp scattering or other explicit momentum-conservation-breaking
mechanisms.

We expect that significant interaction and temperature effects on conductivity will occur at temperatures comparable to or
higher than the chemical potential, i.e., $k_BT\gtrsim\mu$.  The reason for this is that, if the temperature is much lower
than the chemical potential, then scattering processes that can alter the sum of the velocities of the electrons are frozen
out due to the Pauli exclusion principle (a similar argument forms the basis for Landau's Fermi-liquid theory).   We note
that, in Dirac systems, therefore, a fundamental difference exists between $\mu <k_BT$ and $\mu >k_BT$ (where $\mu=0$ is the
Dirac point or the band touching point in our notation), with the higher-temperature regime corresponding to an ``intrinsic''
non-Fermi-liquid-type system where interactions matter nontrivially, while the lower-temperature regime is more like a
standard Fermi-liquid system.  It follows that the pure semimetal with the chemical potential precisely at the Dirac point
is always a non-Fermi liquid and always explicitly violates Kohn's theorem.

We should emphasize that the linear Dirac cone nature of the dispersion is not the ultimate source of this breakdown of
Kohn's theorem, but rather the presence of a filled Fermi sea of negative-energy electrons.  To help illustrate this, we
also investigated the case of materials with quadratic band touchings in their spectra, such as bilayer graphene, and showed
that these two theorems break down in them as well.  The common theme with these is the presence of an ``infinite'' sea of
electrons and the lack of a band gap.  If a band gap were to be opened, then this would freeze out scattering processes
that involve the Fermi sea, thus restoring Kohn's theorem as long as the temperature is much less than the band gap.  This
other method for rescuing these theorems, of course, is relevant only when the chemical potential is close to the band minimum;
otherwise, there is no real difference from the case of a large chemical potential discussed above.  We should also emphasize
that our results rely on the fact that the two bands touch at a point, rather than overlap with each other, with the chemical
potential within the overlap region, as in an ordinary semimetal.  In such a case, we will see no zero-energy electron-hole
excitations of the kind ultimately responsible for the effects described in this work at zero temperature, and thus, Kohn's
theorem will apply to ordinary semimetals at low temperature.

Thus, a gapless semimetal with the chemical potential pinned at the band touching point with a completely filled Fermi sea and a completely
empty Fermi sea (i.e., an intrinsic undoped Dirac-Weyl system) is always a non-Fermi liquid (independent of the energy band dispersion)
in the sense that interactions affect its conductivity even in the absence of disorder.  The result arises simply from the presence
of zero-energy {\it interband} excitations which make the system fundamentally different from a single-band metal with a partially
filled Fermi sea.  If the chemical potential $\mu$ is finite (i.e., away from the Dirac point), the system still behaves as an
intrinsic material as long as the temperature is high enough: $k_BT\gg\mu$.  It is interesting to note that $k_BT>\mu$ is, in some sense,
the classical limit of the system, and the classical limit manifests a strong quantum critical effect of the underlying Dirac point,
whereas the quantum limit, $\mu\gg k_BT$, is benign and behaves as an ordinary metal.  The reason for this apparently puzzling behavior
is physically obvious: Only in the high-temperature limit do the effective low-energy {\it interband} excitations proliferate, leading to
the strange quantum critical behavior involving the violation of Kohn's theorem discussed in the current work.

As for the (apparent) failure of the $f$-sum rule, it can be traced back in part to the presence of the infinite negative-energy
Dirac sea of electrons obtained only in the low-energy effective theory.  This infinite sea is simply an artifact of the low-energy
approximation---an exact theory of such materials with all bands included possesses no such infinite sea.  Due to this, there is only
a finite amount of weight that interactions can redistribute in the $f$ sum.  This is not the case for the low-energy effective
theory; the presence of the infinite Dirac sea means that there is an infinite amount of ``weight'' present.  This is, of course, in
addition to the fact that the double commutator $[[H,\rho(\vec{q})],\rho(-\vec{q})]$ yields an operator with an expectation value
that will depend on the details of the energy distribution of the electrons.  We still expect that, at least at sufficiently low frequencies
(far below the minimum energy of all occupied single-electron levels), phenomena that appear to violate the $f$-sum rule will occur
because the low-energy theories that imply them are still good approximations at such frequencies but that these are simply due to
the redistribution of weight to frequencies higher than those considered within a given experiment.  We expect that such phenomena
should occur under the same conditions as those due to violations of Kohn's theorem, i.e., when the temperature $k_BT\gg\mu$.

\acknowledgements

This work is supported by the Laboratory for Physical Sciences.

\end{document}